# Rejoinder: Fisher Lecture: Dimension Reduction in Regression

R. Dennis Cook

## 1. INTRODUCTION

I am grateful to all of the discussants for their comments which raise a number of important and insightful issues, and add significantly to the breadth of ideas. Following a few introductory comments on the need for a new regression genre that centers on dimension reduction, I turn to the discussants' remarks.

The development in the 1960s and early 1970s of diagnostic methods for regression produced a major shift in regression methodology. When a diagnostic produces compelling evidence of a deficiency in the current model or data it is natural to pursue remedial action, leading to a new model and a new round of diagnostics, proceeding in this way until the required diagnostic checks are passed. By the late 1970s this type of iterative model development paradigm was widely represented in the applied sciences and was formalized in the statistical literature by Box (1979, 1980) and Cook and Weisberg (1982). With the availability of desktop computing starting in the mid-1980s, it is now possible to apply in reasonable time batteries of graphical and numerical diagnostics to many regressions.

Advances in computing and other technologies now allow scientists to routinely formulate regressions in which the number $p$ of predictors is considerably larger than that normally considered in the past. Such large-$p$ regressions necessitate a new type of analysis for at least two reasons. First, the standard iterative paradigm for model development can become untenable when $p$ is large. Recognizing the variety of graphical diagnostics that could be used and the possibility of iteration, a thorough analysis might require assessment of many plots in addition to various numerical diagnostics. Experience has shown that the paradigm can often become imponderable when applied with too many predictors. Second, in some regressions, particularly those associated with high-throughput technologies, the sample size $n$ may be smaller than $p$, leading to operational problems in addition to ponderability difficulties. These issues have caused a shift in the applied sciences toward a different regression genre with the goal of reducing the dimensionality of the vector $\mathbf{X} \in \mathbb{R}^p$ of predictors as a first step in the analysis, effectively raising an old idea to a position of prominence.

Today, dimension reduction is ubiquitous in the applied sciences, represented primarily by principal component methodology. Fifteen years ago I rarely encountered intra-university scientists seeking help with principal component reductions in regression. Such settings no longer seem unusual. The reasons for this are as indicated previously: While I occasionally see problems with $n < p$, more frequently $n$ is several times $p$, while $p$ itself is too large for a full commitment to iterative model development guided by diagnostics. This, in addition to the reasons stated in Section 2 of the article, leads me to conclude that the case for dimension reduction methodology has been made, methodology based on firm parametric foundations with subsequent robust and nonparametric counterparts. Whether the ideas and methodological directions I proposed will meet this goal is less clear, but I am still convinced that they hold promise when $\mathbf{X}$ and $Y$ are jointly distributed.

In contrast to a comment by Christensen, I think parametric dimension reduction is currently as important, if not more important, than other forms, partly because dimension reduction methodology has existed mostly in a world apart from core Fisherian theory, making it difficult to appreciate what could be achieved. For this reason I welcome Christensen's

R. Dennis Cook is Professor, School of Statistics, University of Minnesota, 224 Church Street S. E., Minneapolis, Minnesota 55455, USA e-mail: dennis@stat.umn.edu.

development of connections with multivariate linear model theory.

## 2. APPLICABILITY

According to Christensen, a key issue in the development of models (2), (5), (10) and (13) is whether they are "broadly reasonable." I agree. Moreover, the emerging picture does seem to be one of broad reasonableness for the reasons indicated in the following sections.

### 2.1 First Applications

As mentioned in Section 6.5 of the article, I have used the reductive model (13) on several standard data sets from the literature and nearly always have found $d < p$, and often substantially so. Additionally, I recently applied model (13) to a consulting problem with about 35 predictors and a little less than 300 observations. Again $d \ll p$ with an apparently good fit.

The logo data analyzed by Li and Nachtsheim offer another opportunity to test the proposed methodology. To complement their analysis I first used Grassmann optimization to fit the reductive model (13) with $d = 1$ and the cubic option for $\mathbf{f}_y$, without predictor screening. The log likelihood ratio statistic (cf. Section 6.5) for comparing this fit to the full model has the value $\Lambda_1 = 94.02$ on $r(p-1) = 63$ degrees of freedom, for a nominal $p$-value of 0.007. The same process with other values of $d$ gave $\Lambda_2 = 66.76$, $\Lambda_3 = 65.1$ and $\Lambda_{21} = 0.076$ with corresponding $p$-values 0.26, 0.21 and 0.99. Consequently, I inferred that the sufficient reduction $\mathbf{\Gamma}^T \mathbf{X}$ is two-dimensional. This analysis provides another instance in support of the possibility that the proposed models are broadly reasonable. Grassmann optimization is preferable to evaluating the log likelihood at various combinations of the eigenvectors of $\widehat{\mathbf{\Sigma}}$ or $\widehat{\mathbf{\Sigma}}_{\mathrm{fit}}$, although the latter can sometimes provide a useful approximation.

There are several ways in which an estimated sufficient reduction $\widehat{\mathbf{\Gamma}}^T \mathbf{X}$ could be used to continue an analysis, depending on application-specific requirements. As long as $d$ is sufficiently small, as it is in the logo data, the standard model-diagnostic paradigm might be used to develop a forward model for the regression of $Y$ on $\widehat{\mathbf{\Gamma}}^T \mathbf{X}$. In cases where a full forward model is not essential, we could proceed directly to estimation of the forward mean function based on the relation

$$\mathrm{E}(Y | \mathbf{X} = \mathbf{x}) = \frac{\mathrm{E}\{Y N(\mathbf{x}|Y)\}}{\mathrm{E}\{N(\mathbf{x}|Y)\}},$$

where $N$ is the normal density of $\mathbf{X}|Y$ and the expectations are with respect to the marginal distribution of $Y$. This can be estimated using

$$\widehat{\mathrm{E}}(Y | \mathbf{X} = \mathbf{x}) = \frac{\sum_{i=1}^n y_i \widehat{N}(\mathbf{x}|y_i)}{\sum_{i=1}^n \widehat{N}(\mathbf{x}|y_i)},$$

where $\widehat{N}$ denotes the estimated density obtained by substituting parameter estimates. The estimated mean function allows construction of residuals $y_i - \widehat{\mathrm{E}}(Y|\mathbf{X} = \mathbf{x}_i)$ that can be plotted to check the forward mean function implied by the inverse model. Here the residuals are used as a final diagnostic check, and not necessarily as an integral part of a model building process.

### 2.2 Supervised Principal Components

Li and Nachtsheim developed a fascinating connection between the reductive model (13) and the supervised principal component (SPC) setting of Bair et al. (2006). The latent variable model used by Bair et al. (2006, Section 2.2) and described by Li and Nachtsheim leads to the PFC model (5) or the reductive model (13), depending on the restrictions placed on the covariance matrix of the errors ($\varepsilon, \epsilon_1, \ldots, \epsilon_p$). However, Bair et al. did not pose or use (5) or (13) as their basis for estimation, but instead used principal components. Thus, the method of SPC's is more in line with the PC model (2). Conformably partition

$$\mathbf{X} = \begin{pmatrix} \mathbf{X}_1 \\ \mathbf{X}_2 \end{pmatrix} \quad \text{and} \quad \mathbf{\Gamma} = \begin{pmatrix} \mathbf{\Gamma}_1 \\ \mathbf{\Gamma}_2 \end{pmatrix},$$

and assume that $\mathbf{\Gamma}_2 = 0$. Then under the PC model (2), the sufficient reduction is just $\mathbf{\Gamma}_1^T \mathbf{X}_1$ and the maximum likelihood estimator of span($\mathbf{\Gamma}_1$) is the span of the first $d$ eigenvectors of the sample version of Var($\mathbf{X}_1$). This leads to exactly the analysis suggested by Bair et al.: Use initial screening to identify $\mathbf{X}_2$ and then compute the first (or first few) principal components for $\mathbf{X}_1$. I expect that PFC reductions based on model (5) or sufficient reductions based on model (13) can do much better than SPC's, particularly if prior screening is based on an information criterion like AIC or BIC applied in the context of the inverse model.

This expectation is supported by the logo data, since there is only a very weak relationship between the first two principal components and the estimated sufficient reduction based on model (13). For instance, the value of $R^2$ from the linear regression of the first principal component on $\widehat{\mathbf{\Gamma}}^T \mathbf{X}$ is only 0.11.

The upper left plot in Li and Nachtsheim's Figure 1 was computed using the first *principal standardized component*, the first principal component computed after standardizing each predictor to have a marginal sample standard deviation of 1 (personal communication). The value of $R^2$ from the linear regression of the first principal standardized component on $\widehat{\boldsymbol{\Gamma}}^T \mathbf{X}$ is 0.60. In this example the first principal standardized component outperformed the first principal component, although this need not always be so (cf. Section 7.3). The general points in my discussion of predictor standardization were captured nicely by Christensen's summary: Standardization is necessary, unless the regression can be described reasonably by the PC model (2) or the PFC model (5), but the common practice of marginal predictor scaling is not sufficient, and may even be counterproductive.

On balance, SPC's are based on relatively weak supervision, since the response is used only in the screening phase. Models (5) and (13) allow more complete supervision, whether used with prior screening or not.

### 2.3 Variance Reduction versus Bias

Consider a regression in which model (13) holds with $d = p$, which is equivalent to model (17), and that no version of model (13) holds strictly with $d < p$. In that case, fitting (13) with $d < p$ will result in bias along with a reduction in variance. It is conjectured that often the reduction in variance will outweigh the increase in bias, resulting in a reduction in mean squared error. In other words, there may be reason to pursue models like (13) with $d < p$ even when they are "incorrect."

### 2.4 Partial Least Squares

Like Christensen, I have had misgivings about partial least squares and found its apparent popularity in some areas, particularly chemometrics, to be a bit curious. However, the developments in the article are causing me to reconsider. In Section 7.4 of the article I developed a connection between OLS and the inverse model (17). Here I establish a connection with partial least squares via the population relationship between OLS and model (13) with $d < p$.

For notational convenience, let $\mathbf{M} = \boldsymbol{\Omega}^2 + \boldsymbol{\beta} \cdot \text{Var}(\mathbf{f}_Y) \boldsymbol{\beta}^T$ and $\mathbf{C} = \text{Cov}(\mathbf{X}, Y)$. Suppose that we wish to estimate the population OLS coefficient vector $\mathbf{B} = \boldsymbol{\Sigma}^{-1} \text{Cov}(\mathbf{X}, Y)$. From Proposition 4, $\boldsymbol{\Sigma} = \boldsymbol{\Gamma}_0 \boldsymbol{\Omega}_0^2 \boldsymbol{\Gamma}_0^T + \boldsymbol{\Gamma} \mathbf{M} \boldsymbol{\Gamma}^T$, and thus $\mathbf{M}^{-1} = (\boldsymbol{\Gamma}^T \boldsymbol{\Sigma} \boldsymbol{\Gamma})^{-1}$ and $\boldsymbol{\Sigma}^{-1} = \boldsymbol{\Gamma}_0 (\boldsymbol{\Omega}_0^2)^{-1} \boldsymbol{\Gamma}_0^T + \boldsymbol{\Gamma} \mathbf{M}^{-1} \boldsymbol{\Gamma}^T$. Now,

$$\begin{aligned}
\mathbf{B} &= \boldsymbol{\Sigma}^{-1} \boldsymbol{\Gamma} \boldsymbol{\beta} \, \text{Cov}(\mathbf{f}_Y, Y) \\
&= \boldsymbol{\Gamma} (\boldsymbol{\Gamma}^T \boldsymbol{\Sigma} \boldsymbol{\Gamma})^{-1} \boldsymbol{\beta} \, \text{Cov}(\mathbf{f}_Y, Y) \\
&= P_{\boldsymbol{\Gamma}(\boldsymbol{\Sigma})} \mathbf{B} \\
&= \boldsymbol{\Gamma} (\boldsymbol{\Gamma}^T \boldsymbol{\Sigma} \boldsymbol{\Gamma})^{-1} \text{Cov}(\mathbf{X}, Y).
\end{aligned}$$

This says that $\mathbf{B} \in \mathcal{S}_{\boldsymbol{\Gamma}}$ and that we can construct a marginal estimator of $\mathbf{B}$ by projecting the usual moment estimate $\widehat{\mathbf{B}}$ onto $\widehat{\mathcal{S}}_{\boldsymbol{\Gamma}}$ relative to the $\widehat{\boldsymbol{\Sigma}}$ inner product.

The population version of the partial least squares coefficients follows this same pattern $\mathbf{B}_{\text{pls}} = P_{\mathbf{K}(\boldsymbol{\Sigma})} \mathbf{B}$ (Helland, 1992; Helland and Almøy, 1994; Naik and Tsai, 2000), except that $\mathcal{S}_{\boldsymbol{\Gamma}}$ is replaced by the cyclic subspace $\mathcal{S}_{\mathbf{K}}$ spanned by $\mathbf{K} = (\mathbf{C}, \boldsymbol{\Sigma} \mathbf{C}, \boldsymbol{\Sigma}^2 \mathbf{C}, \ldots, \boldsymbol{\Sigma}^{q-1} \mathbf{C})$ for some integer $q$ that is often chosen by cross-validation in practice. Assume that $\mathbf{C}$ can be written as a linear combination of at most $q$ eigenvectors of $\boldsymbol{\Sigma}$. Then $\mathbf{B} \in \mathcal{S}_{\mathbf{K}}$ (Naik and Tsai, 2000) and consequently the marginal estimator of $\mathbf{B}$ from model (13) and the PLS estimator have the same basic form, but differ on the method of estimating an upper bound—$\mathcal{S}_{\mathbf{K}}$ or $\mathcal{S}_{\boldsymbol{\Gamma}}$—for span($\mathbf{B}$). Partial least squares estimates can be computed without inversion of $\widehat{\boldsymbol{\Sigma}}$, which seems to be one of their attractions. In short, the proposed methods inherit support from the apparent reasonableness of partial least squares in some contexts.

Principal components, principal fitted components, partial least squares and reductive methods based on the inverse model (13) all attempt to make use of information from the marginal distribution of $\mathbf{X}$ when inferring about $Y|\mathbf{X}$. This distinguishes them from methods like OLS, SIR, RMAVE and the lasso that apparently do not consider such information. The simulation results in the article show that substantial gains over forward methods are possible when $\mathbf{X}$ contains information on $Y|\mathbf{X}$ via $\boldsymbol{\Sigma}$. It seems fair to conclude that the models considered in the article will be broadly reasonable at least within the class of regressions where $\mathbf{X}$ is informative about $Y|\mathbf{X}$.

## 3. NONCONSTANT VAR($\mathbf{X}|Y$)

The development throughout the article was based on the assumption that $\text{Var}(\mathbf{X}|Y)$ is constant. If $\text{Var}(\mathbf{X}|Y)$ is not constant, then the reductions described here may no longer be sufficient, although they will still be functions of sufficient reductions.

Li and Nachtsheim describe an extreme case of this where the forward mean function $E(Y|\mathbf{X})$ is quadratic in one of the predictors without a linear trend. For example, when $E(Y|\mathbf{X}) = X_1^2$ and $X_1$ is symmetrically distributed about 0, $E(X_1|Y) = 0$ while $\text{Var}(X_1|Y)$ will be nonconstant. The potential for this kind of setting can be detected by using a numerical diagnostic for heteroscedasticity in the context of fitting a simple linear model to $X_j|Y$, $j = 1, \ldots, p$.

B. Li has taken an interesting first step in the study of models that allow for nonconstant $\text{Var}(\mathbf{X}|Y)$. His Theorem 2.2 shows that we can construct sufficient reductions for both the conditional mean and the conditional variance, and thus cover settings like that described by L. Li and Nachtsheim. In the context of his model (5), the methods of this article should be useful for estimating $\text{span}(\mathbf{\Gamma}_1)$, but they will not be able to identify the part of $\text{span}(\mathbf{\Gamma}_2)$ that is not contained in $\text{span}(\mathbf{\Gamma}_1)$. However, even if $\text{span}(\mathbf{\Gamma}_1) = \text{span}(\mathbf{\Gamma}_2)$, there should be efficiency gains by considering models like B. Li's (5). These gains are nicely illustrated by Li's examples. Additionally, analyses based on heteroscedastic inverse models will likely encounter inference issues not considered in the article. For instance, it may be of interest to test if $\text{span}(\mathbf{\Gamma}_2) \subseteq \text{span}(\mathbf{\Gamma}_1)$ or vice versa.

## 4. PRINCIPAL COMPONENTS

B. Li presented an intriguing explanation for why the response tends to have a higher correlation with the first principal component than any other component, but concluding that dimension reduction via $\mathbf{X}|Y$ should offer substantial gains. I expect that his Conjecture 1.1 is correct and that it offers a partial explanation for the popularity of reduction to a few leading principal components. We may be able to gain additional insights into the situation reasoning as follows, using Li's notation and model. Recall that $X$ is $N(0, \Sigma)$, $Y = \beta^T X + \epsilon$ and $X \perp\!\!\!\perp \epsilon$.

Hold fixed $\beta$ with $\|\beta\| = 1$ and $\Sigma$ with eigenvalues $\lambda_1 > \cdots > \lambda_p > 0$, and assume that the errors $\epsilon$ are normally distributed with mean 0 and variance $\sigma_\epsilon^2$. Then the squared correlation coefficient between $Y$ and the first principal component $v_1^T X$ can be expressed as

$$\rho_1^2 = \frac{(\beta^T v_1)^2 \lambda_1}{\sigma_\epsilon^2 + \sum_{j=1}^p (\beta^T v_j)^2 \lambda_j}.$$

If the eigenvalues $\lambda_j$ are roughly the same, then the magnitude of $\rho_1$ is controlled by $\sigma_\epsilon^2$ and the angles between $\beta$ and the eigenvectors $v_j$. However, if $\beta^T v_1 \neq 0$ and $\lambda_1$ is sufficiently larger than both $\lambda_2$ and $\sigma_\epsilon^2$, then $\rho_1^2$ will be close to 1. This might be taken to suggest that reduction to the first principal component is desirable, particularly when $\lambda_1 \gg \lambda_2$.

However, $Y|v_1^T X$ is normally distributed with mean $E(Y|v_1^T X) = \beta^T v_1 v_1^T X$ and variance

$$\text{Var}(Y|v_1^T X) = \sigma_\epsilon^2 + \sum_{j=2}^p (\beta^T v_j)^2 \lambda_j.$$

This distribution does not depend on the value of $\lambda_1$. Provided $\beta \neq v_1$, $\text{Var}(Y|v_1^T X) > \text{Var}(Y|\beta^T X)$, reflecting the fact that $v_1^T X$ is not a sufficient reduction and consequently that conditioning on $v_1^T X$ does not exhaust the information that $X$ contains about $Y$, regardless of the magnitude of $\rho_1$.

## 5. BINARY PREDICTORS

The main thrust of my article is on normal inverse models, but I also suggested how the ideas could be applied with conditional predictors $\mathbf{X}|Y$ from other families. I was particularly pleased to see that Li and Nachtsheim implemented an algorithm for regressions with all binary predictors, but was simultaneously a bit disappointed to see that it did not work out as crisply as expected. Their suggestion of a majorization strategy to resolve the optimization issues is excellent and will likely produce a stable and practically useful algorithm. Majorization may not be essential for similar algorithms developed for principal fitted components ($\boldsymbol{\nu}_y = \boldsymbol{\beta} \mathbf{f}_y$), although it might still improve performance.

## ACKNOWLEDGMENTS

The author is grateful to the Executive Editor, Ed George, for arranging this discussion and to Liliana Forzani for helpful comments. Research for this discussion was supported in part by National Science Foundation Grant DMS-04-05360.